\documentclass[conference]{IEEEtran}
\IEEEoverridecommandlockouts

\usepackage{cite}

\usepackage{orcidlink}

\usepackage{multirow}
\usepackage{graphicx} 
\usepackage{amsmath,amssymb,amsfonts}
\usepackage{algorithmic}
\usepackage{graphicx}
\usepackage{textcomp}
\usepackage{float}
\usepackage{xcolor}
\def\BibTeX{{\rm B\kern-.05em{\sc i\kern-.025em b}\kern-.08em
    T\kern-.1667em\lower.7ex\hbox{E}\kern-.125emX}}
\begin{document}

\title{Model-Driven Quantum Code Generation Using Large Language Models and Retrieval-Augmented Generation
}

\author{\IEEEauthorblockN{Nazanin Siavash\,\hypersetup{pdfborder={0 0 0}}\orcidlink{0009-0000-4177-0632}}
\IEEEauthorblockA{\textit{Department of Computer Science} \\
\textit{University of Colorado Colorado Springs (UCCS)}\\
Colorado Springs, United States \\
nsiavash@uccs.edu}
\and
\IEEEauthorblockN{Armin Moin\,\hypersetup{pdfborder={0 0 0}}\orcidlink{0000-0002-8484-7836}}
\IEEEauthorblockA{\textit{Department of Computer Science} \\
\textit{University of Colorado Colorado Springs (UCCS)}\\
Colorado Springs, United States \\
amoin@uccs.edu}
}

\maketitle

\begin{abstract}
This paper introduces a novel research direction for model-to-text/code transformations by leveraging Large Language Models (LLMs) that can be enhanced with Retrieval-Augmented Generation (RAG) pipelines. The focus is on quantum and hybrid quantum-classical software systems, where model-driven approaches can help reduce the costs and mitigate the risks associated with the heterogeneous platform landscape and lack of developers' skills. We validate one of the proposed ideas regarding generating code out of UML model instances of software systems. This Python code uses a well-established library, called Qiskit, to execute on gate-based or circuit-based quantum computers. The RAG pipeline that we deploy incorporates sample Qiskit code from public GitHub repositories. Experimental results show that well-engineered prompts can improve CodeBLEU scores by up to a factor of four, yielding more accurate and consistent quantum code. However, the proposed research direction can go beyond this through further investigation in the future by conducting experiments to address our other research questions and ideas proposed here, such as deploying software system model instances as the source of information in the RAG pipelines, or deploying LLMs for code-to-code transformations, for instance, for transpilation use cases.
\end{abstract}


\section{Introduction} \label{Introduction}
Large Language Models (LLMs) have recently been deployed in various aspects of Software Engineering (SE) \cite{Hou+2024}, including but not limited to the Model-Driven Software Engineering (MDSE) paradigm \cite{MosthafWasowski2024}. For instance, when it comes to bridging the divide between domain experts and MDE tools, generative AI seems to be promising \cite{Kulkarni+2023}. Within the realm of MDSE, LLMs have been applied to various tasks, including model management, code generation, user assistance and training, automation of modeling tasks, as well as error detection and correction \cite{DiRocco+2025}.

Moreover, MDSE has proven useful in complex vertical domains and use-case scenarios in which heterogeneous platforms and technology stacks need to be supported. Quantum Software Engineering (QSE) deals with such a landscape. Additionally, today's software developers typically lack technical skills in these platforms, for example, regarding their models of computation, hardware architecture, Software Development Kits (SDKs), libraries, constraints, and optimization techniques. For instance, Gemeinhardt et al. \cite{Gemeinhardt+2024} proposed a novel approach to enable model-driven composition-based quantum circuit design and implementation. Also, Jiménez-Navajas et al. \cite{Jimenez+2025} deployed the Epsilon Generation Language to generate Python code that works with Qiskit for quantum and hybrid quantum-classical applications. We used the model instances provided by them in our experimental study.

The key contribution of this paper is proposing a new research direction defined by deploying LLMs, such as the state-of-the-art Generative Pre-trained Transformer (GPT) models provided by OpenAI (e.g., GPT-4o), as a text generation engine (e.g., to automatically generate source code and documentation) in MDSE. In this paper, we concentrate on generating software code that should execute on quantum computers. The code will be generated by the LLM out of the software model instances, modeling the behavior and structure of the respective software systems. Furthermore, rather than relying exclusively on prompt-based LLM generation, which often suffers from limited context-awareness, our approach integrates a Retrieval-Augmented Generation (RAG) pipeline. This enhancement grounds the LLM outputs in relevant domain-specific artifacts, thereby improving both the accuracy and consistency of the generated quantum code. RAG is a well-established, state-of-the-art concept to address the so-called LLM hallucination problem, which is abundant~\cite{Fan+2024}.

The structure of this paper is as follows. Section \ref{Background} provides background information. Section \ref{related-work} reviews related work in these areas. In Section \ref{proposed-approach}, we describe our proposed research direction. Moreover, Section \ref{preliminary-experimental-result} presents our preliminary experimental results. Finally, Section \ref{conclusion-and-future-work} concludes the paper and outlines the future work.

\section{Background} \label{Background}
In the following, we briefly provide some background information on LLM-enabled code generation and code transpilation.


\subsection{LLM-enabled Code Generation} \label{SS:QCG}
Classic code generation approaches in MDSE have deployed template-based and rule-based model transformation techniques. However, with recent advances in machine learning, LLM-based approaches have gained significant traction \cite{LeAndrzejak2024,FanGokkaya+2023}. These approaches have notably improved the accuracy, coherence, and fluency of code synthesis and completion \cite{Chen+2024}. A variety of LLMs are available, and the choice depends on the specific task requirements and computational resources. Common examples include GPT-4, LaMDA, Claude, and LLaMA 2.

\subsection{Code Transpilation} \label{SS:QCT}
Transpilation refers to the process of converting source code from one programming language to another, with the goal of maintaining the original program’s intended semantics and functionality \cite{Nitin+2025}. Traditionally, this transformation has been achieved through rule-based or template-driven approaches that depend on explicitly defined syntactic and semantic mappings, often requiring substantial manual effort and deep domain expertise \cite{Roziere+2020}. In recent years, however, there has been a paradigm shift toward the utilization of LLMs for this purpose \cite{Yang+2024,Pan+2024,SiavashMoin2025}. Trained on extensive corpora of programming data, these advanced generative machine learning models are capable of producing target-language code that adheres to established coding conventions and standards \cite{Nitin+2025}. Consequently, LLM-based approaches have demonstrated considerable effectiveness in maintaining semantic fidelity during the transpilation process.

\section{Related Work} \label{related-work}
Below, we briefly review the recent developments in automated source code generation and transpilation.

\subsection{Rule-based Code Generation}
Traditional code generation and transpilation techniques often rely on explicitly defined transformation rules. Various domain-specific modeling tools based on scientific approaches have been proposed, for example, MontiThings \cite{Kirchhof+2022a} for the Internet of Things (IoT), MontiAnna \cite{Kirchhof+2022b} for machine learning, and ML-Quadrat \cite{Moin+2022a} for machine learning-enabled IoT services. Additionally, Xue et al. \cite{Xue+2020} presented the PAR platform, a formal model-driven engineering framework incorporating languages and transformation rules to automatically derive executable programs from high-level specifications. Also, Ling et al. \cite{Ling+2022} tackled (code-to-code) transpilation with a focus on API safety. 

The idea of deploying MDE for QSE \cite{Gemeinhardt+2021} and code generation for various quantum AI \cite{Moin+2022-MDE4QAI,Moin+2023-QFL} has also been around over the past few years. Jiménez-Navajas et al. \cite{Jimenez+2025} enabled generating quantum code from extended UML models using model-to-text transformations written in the Epsilon Generation Language. Their approach supported generating Python code that could work with the Qiskit library for hybrid quantum-classical systems. It was validated through multiple case studies, confirming its practical applicability. We use their model instances for our preliminary evaluation.

Further, Gemeinhardt et al. \cite{Gemeinhardt+2024} developed a domain-specific modeling language to facilitate quantum circuit design at a higher level of abstraction. Their toolchain enabled the specification of composite quantum operations and automated the generation of corresponding executable code, demonstrated through applications such as Quantum Counting and Quantum Approximate Optimization Algorithm (QAOA).

\subsection{LLM-enabled Code Generation}
With the increasing complexity of software systems and the limitations of rule-based methods, recent work has shifted toward leveraging LLMs for code generation and transpilation. These models eliminate the need for handcrafted transformation rules by learning patterns from large code corpora. Zheng et al. \cite{Zheng+2023} introduced CodeGeeX, a 13-billion-parameter multilingual model trained across 23 programming languages. It demonstrated strong performance in both code synthesis and transpilation tasks, outperforming comparable models of the same scale. Furthermore, Le and Andrzejak \cite{LeAndrzejak2024} proposed a novel approach called One-shot Correction, which decomposed user queries into sub-tasks, retrieved or generated relevant code snippets using stored user feedback or a Natural Language-to-code generator, and assembled them into final code.

While LLMs have been extensively studied for classical code generation, there is a notable lack of research applying these techniques to quantum software development. Most recently, Henderson et al. \cite{Henderson+2025} studied the use of LLMs for quantum programming. However, their experiments have not been with the state-of-the-art LLMs, such as GPT-4o or GPT-4.1, but with the previous OpenAI LLM version, GPT-4. Furthermore, they did not investigate the deployment of RAG pipelines.



\section{Proposed Research Direction} \label{proposed-approach}
\begin{figure*}[ht]
\centering
\includegraphics[width=2\columnwidth]{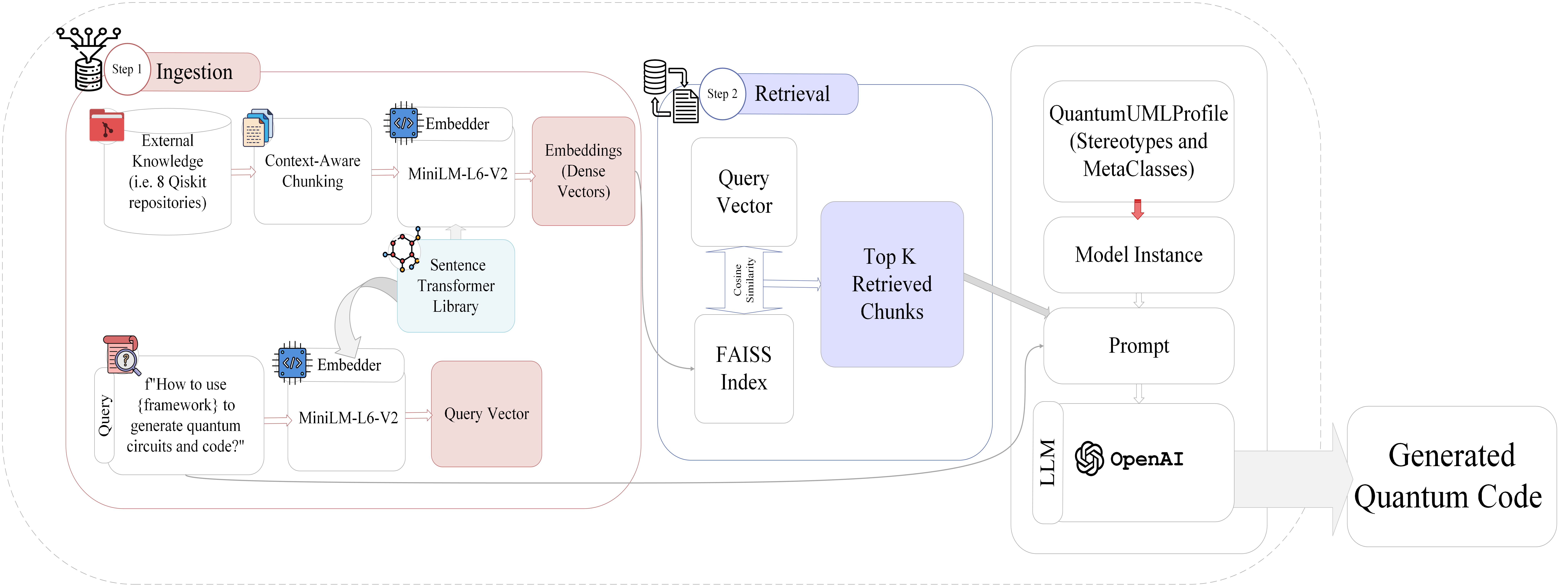}
\caption{The overall architecture of our current RAG pipeline configuration}
\label{fig:QCTG}
\end{figure*}

The proposed research direction, which is based on our novel vision, comprises leveraging LLMs enhanced by RAG pipelines to mitigate their well-known hallucination problem and increase their reliability. Hence, we propose the following Research Questions (RQ) for various configurations that we can imagine at this time for the realization of this vision: RQ 1. Can an LLM receive a design model instance of a software system or component in a general-purpose modeling language, such as UML, as input, and generate the implementation corresponding to that model in a general-purpose programming language as output in an effective and efficient manner? RQ2. Can a RAG pipeline fed by sample code snippets in the target programming language help enhance the effectiveness and/or efficiency of the code generation in RQ1? RQ3. Can prompt engineering techniques enhance the performance of the LLM? RQ4. Can we support a domain-specific modeling language in RQ1 and RQ2, and provide the meta-model of the language as additional context in the RAG pipeline? RQ5. Can we change the LLM input and the configuration of the RAG pipeline in RQ2 (or RQ4), such that the LLM input is the software requirements specification in natural language, and the corresponding model instance is deployed in the RAG pipeline providing additional context? RQ5. Can we use the RAG setup proposed in RQ4 (i.e., the model instance as the RAG-enabled context to augment the prompt/query for the purpose of code-to-code transformation, such as in transpilation (i.e., translation of the code from one variant to another one, e.g., from one programming language to another one)? Our emerging results presented in Section \ref{preliminary-experimental-result} pertain to a basic prototype realized with the aim of addressing RQ1, RQ2, and RQ3 among the new ideas presented above. The remaining RQs, RQ4–RQ6, are outlined as directions for future work. One of the key novelties of the proposed approach is its focus on quantum and hybrid quantum-classical software applications rather than pure classical ones.

\subsection{LLM-enabled Model-Driven Quantum Code Generation}

We enable automated Python source code generation via the OpenAI's LLMs, specifically GPT-4o. The input prompt given to the LLM entails the textual model instance that conforms to a meta-model. The output of the LLM is the generated code that can work with IBM's Qiskit quantum software library.

\subsection{Context-aware Model-Driven Quantum Code Generation via a RAG pipeline}
To overcome the LLM limitations that cause occasional hallucinations, a RAG approach is employed \cite{Rani+2024}. This method integrates a retrieval system with a generative model, enabling dynamic access to external knowledge during the generation process. Specifically, the RAG architecture includes two main parts: A retriever that identifies relevant documents from a knowledge base, and a generator that uses this retrieved information to inform the final output. The overall architecture of our RAG pipeline, in its present configuration, is illustrated in Figure \ref{fig:QCTG}.

\subsection{Prompt Engineering}
Designing effective prompts is a key step in improving the performance of LLMs on specific tasks \cite{Ye+2023}. This process, known as \textit{prompt engineering}, involves crafting and refining the input provided to LLMs, such as GPT-4o deployed here, to clearly express the user's intent \cite{Pawar+2024,ClarisóCabot2023}. We examine two types of prompts. The first one is a more general prompt that simply instructs the LLM to generate quantum code based on the quantum circuit defined in the input model instance. This version provides minimal guidance and relies on the LLM’s default understanding. The second prompt variant is more specific, incorporating detailed implementation requirements, such as quantum gate mapping strategies and constraints on syntax and gate behavior. By comparing the outputs from these two prompting strategies, we aim to evaluate how varying levels of prompt-specificity influence the quality and correctness of the generated quantum code.

\section{Preliminary Experimental Results}\label{preliminary-experimental-result}

\subsection{Experimental Data}
We carry out our experimental study with the model instances presented in Jiménez-Navajas et al. \cite{Jimenez+2025}, which conform to a Unified Modeling Language (UML) Profile proposed by P{\'e}rez-Castillo and Piattini \cite{Perez+2022}. 

\subsection{Evaluation Metrics}
To assess the quality of the quantum code generated from UML models, we adopt a structured evaluation approach based on Precision, Recall, and F-measure. These metrics are calculated by comparing the elements present in the UML model (expected elements) with those identified in the generated code (generated elements). The comparison methodology follows the element-wise mapping defined in the research paper by Jiménez-Navajas et al.  \cite{Jimenez+2025}. 


\subsubsection{Precision}
Precision evaluates the ratio of correctly generated elements (relevant elements) to the total number of generated elements.

\begin{equation}
\text{Precision} = \frac{\text{Relevant}}{\text{Relevant} + \text{Irrelevant}}
\label{eq:precision}
\end{equation}
Where Relevant and Irrelevant denote the number of relevant and irrelevant elements, respectively. Therefore, precision measures the correctness of what was generated.

\subsubsection{Recall}
Recall quantifies the proportion of relevant elements that were successfully generated, out of all elements that were expected based on the UML model.
\begin{equation}
\text{Recall} = \frac{\text{Relevant}}{\text{Relevant} + \text{Missing}}
\label{eq:recall}
\end{equation}
This allows us to measure how complete the code generation was in covering the intended UML semantics.

\subsubsection{F-measure}
F-measure is the harmonic mean of Precision and Recall. It balances the trade-off between Precision (correctness) and Recall (completeness). It provides a single score to evaluate the overall effectiveness of the generated code.
\begin{equation}
\text{F-Measure} = \frac{2 \cdot \text{Precision} \cdot \text{Recall}}{\text{Precision} + \text{Recall}}
\label{eq:fmeasure}
\end{equation}

Furthermore, to evaluate the quality of the generated code beyond surface-level similarity, this study adopts a machine translation-inspired metric, known as Bilingual Evaluation Understudy (BLEU), which is specifically tailored for programming languages, namely CodeBLEU. 

\subsubsection{CodeBLEU}
Unlike conventional BLEU \cite{Papineni+2002}, CodeBLEU incorporates additional dimensions to assess both syntactic and semantic correctness, providing a more comprehensive evaluation of code generation outputs. CodeBLEU is formulated as a weighted combination of four components \cite{Ren+2020}:
\begin{equation}
\text{CodeBLEU} = \alpha \cdot \text{BLEU} + \beta \cdot \text{BLEU}_{\text{weight}} + \gamma \cdot \text{Match}_{\text{ast}} + \delta \cdot \text{Match}_{\text{df}}
\label{eq-codebleu}
\end{equation}
In Equation \ref{eq-codebleu}, BLEU represents the standard n-gram matching score as defined by Papineni et al. \cite{Papineni+2002}, BLEU$_{\text{weight}}$ refers to a weighted n-gram matching, where tokens from the generated (hypothesis) and reference code are compared by assigning different importance levels to each token, Match$_{\text{ast}}$ captures syntactic similarity by analyzing the correspondence between the Abstract Syntax Trees (ASTs) of the generated and reference code, and Match$_{\text{df}}$ evaluates semantic similarity by analyzing how data flows within the generated code in comparison to the provided reference code.

\subsection{Evaluation Results}
\begin{figure}[ht!]
\centering
\includegraphics[width=0.45\textwidth]{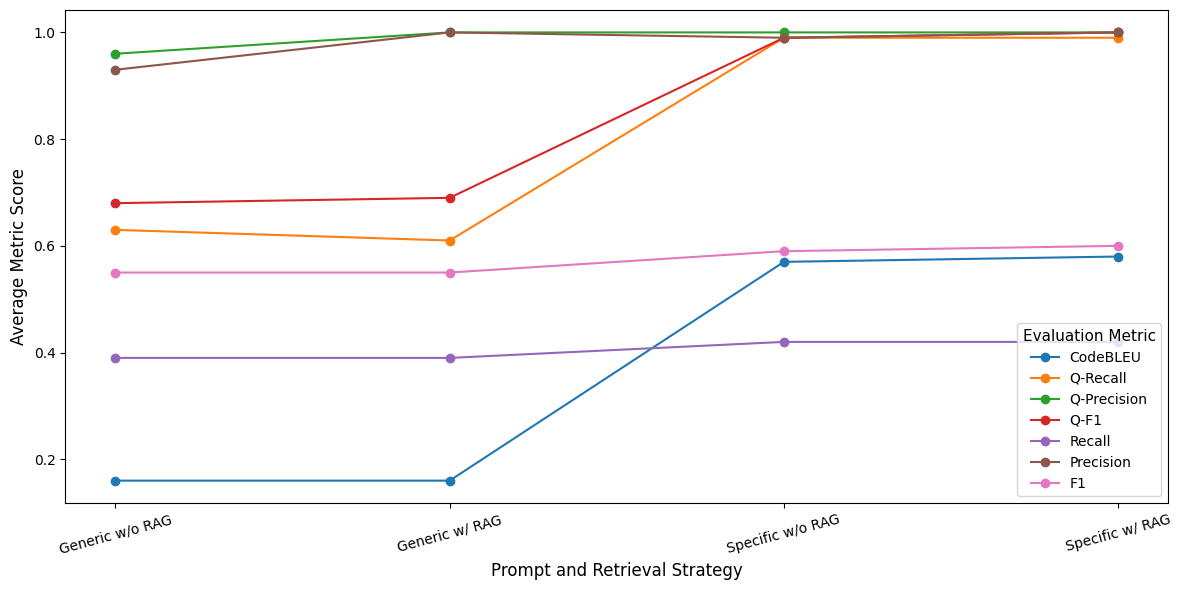}\caption{Effect of prompt specificity and RAG on quantum code generation metrics
}
\label{fig:experimentalresult}
\end{figure}

\begin{figure*}[ht]
\centering
\includegraphics[width=0.7\textwidth]{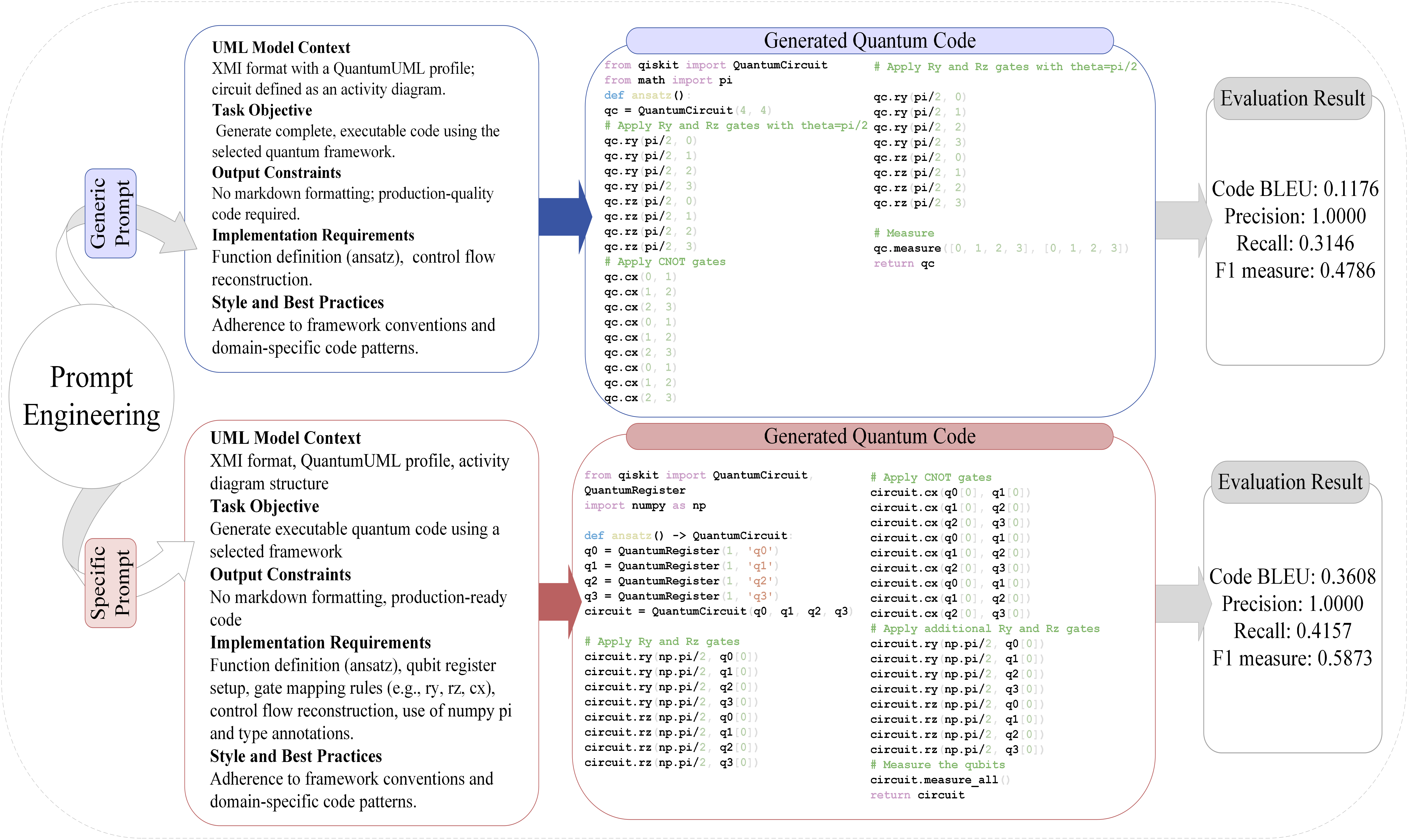}\caption{Evaluation results for generic and specific prompts on model instance 4 of Jiménez-Navajas et al. \cite{Jimenez+2025}}
\label{fig:specific-vs-generic-prompt}
\end{figure*}

We conduct experiments on all seven model instances presented by Jiménez-Navajas et al. \cite{Jimenez+2025}. However, due to space limitations, here we present detailed results only for model instance 1. This one has demonstrated the best overall performance across all evaluation metrics. To complement this, we provide a comparative summary across all of the selected model instances. The best average CodeBLEU score is observed for model instance 3 using a specific prompt without RAG, achieving 0.63. In contrast, the lowest CodeBLEU score is 0.10, reported for model instance 7 with a generic prompt and RAG. For Q-Precision (average of Precision scores for quantum gates and quantum partitions), the maximum score of 1.00 is achieved by multiple configurations, including: Model instances 2 and 4 with specific prompts (both with and without RAG), model instance 7 with a specific prompt and RAG, and model instance 1 with generic and specific prompts, both with and without RAG. The lowest Q-Precision is observed in model instance 3 using a generic prompt, with and without RAG, both scoring 0.30. Regarding Q-Recall (average of Recall scores for quantum gates and quantum partitions), model instance 6 achieves the highest score of 1.00 using a specific prompt, both with and without RAG. The lowest Q-Recall, 0.28, is seen in model instance 7 using a generic prompt with RAG. For Q-F-measure (average of F-measure scores for quantum gates and quantum partitions), the highest value of 1.00 is attained by model instance 1 with a specific prompt and RAG. The lowest Q-F-measure is 0.36, recorded for model instance 7 using a generic prompt with RAG.

Detailed evaluation results for model instance 1 under four different experimental settings are provided in Tables \ref{tab:model1_airbus-generic-withoutRAG}, \ref{tab:model1_airbus-generic-withRAG}, \ref{tab:model1_airbus-specific-withoutRAG}, and \ref{tab:model1_airbus-specific-withRAG}.
Table \ref{tab:model1_airbus-generic-withoutRAG} presents the evaluation results of the standalone LLM-based quantum code generation using a generic prompt.
Table \ref{tab:model1_airbus-generic-withRAG} shows the results under the same generic prompt setting but with external context provided through a RAG pipeline that incorporates eight Qiskit GitHub repositories. This external context does not significantly improve performance across most metrics, hence not validating RQ2 so far. While Q-Precision and Precision show a slight increase (by approximately 0.04–0.07), other metrics remain largely unaffected. This indicates that the selected Qiskit repositories do not offer sufficiently relevant context to enhance quantum code generation from UML model instances. This insight opens new directions for exploring more domain-specific and structured external sources, and/or focusing on the other ideas proposed in Section \ref{proposed-approach}.
Table \ref{tab:model1_airbus-specific-withoutRAG} reports the evaluation results using a more specific prompt, which includes implementation constraints, such as gate mapping instructions and syntax rules. This modification leads to a notable improvement in all metrics, thus validating RQ3. Specifically, the average CodeBLEU score increases from 0.16 to 0.57, Q-Recall improves from 0.63 to 0.99, Q-Precision from 0.96 to 1.00, and Q-F-measure from 0.68 to 0.99. Additionally, Recall improves from 0.39 to 0.42, Precision from 0.93 to 0.99, and F-measure from 0.55 to 0.59. These results confirm that a well-engineered, specific prompt can significantly enhance the LLM's performance in generating the target quantum code.

Table \ref{tab:model1_airbus-specific-withRAG} combines the specific prompt from Table \ref{tab:model1_airbus-specific-withoutRAG} with the same external context used in Table \ref{tab:model1_airbus-generic-withRAG} via the RAG pipeline. The results, however, do not show a significant improvement over Table \ref{tab:model1_airbus-specific-withoutRAG}, reinforcing the earlier observation that the current choice of external repositories and/or this specific RAG configuration provide rather minimal additional value. It is anticipated that more substantial gains may be realized through the incorporation of more relevant external context and/or the adoption of an alternative RAG configuration \cite{Rani+2024, SaberiFard2024}. Comparing Table \ref{tab:model1_airbus-generic-withoutRAG} with Table \ref{tab:model1_airbus-specific-withoutRAG}, and Table \ref{tab:model1_airbus-generic-withRAG} with Table \ref{tab:model1_airbus-specific-withRAG} further highlight that prompt-engineering has a far greater impact on the LLM's code generation performance than the provided external context through the RAG pipeline in the current setup. Figure \ref{fig:experimentalresult} provides a visual representation of this comparative analysis.

For quantum-specific metrics, we achieve near-perfect results with Q-Precision, Q-Recall, and Q-F-measure all at or near 1.0, indicating high semantic correctness and completeness of the generated code. On the other hand, for overall Precision, Recall, and F-measure (including non-quantum elements), we achieve scores of 1.0, 0.42, and 0.60, respectively. While these fall short of the respective scores reported by Jiménez-Navajas et al. \cite{Jimenez+2025} (0.973, 1.0, 0.986), this is expected since our generation focuses solely on the quantum circuit portion of the UML model, rather than the entire model.

\begin{table*}[ht!]
\centering
\caption{Performance metrics across 10 runs for model instance 1 of Jiménez-Navajas et al. \cite{Jimenez+2025} using a generic prompt without RAG}
\label{tab:model1_airbus-generic-withoutRAG}
\begin{tabular}{|c|c|c|c|c|c|c|c|c|c|}
\hline
\multirow{11}{*}{\rotatebox{90}{\textbf{Model Instance 1}}}& \textbf{RUN} & \textbf{CodeBLEU} & \textbf{Q-Recall}& \textbf{Q-Precision}& \textbf{Q-F-measure} & \textbf{Recall} & \textbf{Precision}& \textbf{F-measure}\\
\cline{2-9}
& 1  & 0.16& 0.63& 0.99& 0.69&0.39&0.97&0.56\\
\cline{2-9}
& 2  & 0.15& 0.63& 1.00& 0.70&0.39&1.00&0.56     \\
\cline{2-9}
& 3  & 0.16& 0.63& 0.99& 0.69&0.39&0.97&0.56\\
\cline{2-9}
& 4  & 0.15& 0.63& 0.99& 0.69&0.39&0.97&0.56\\
\cline{2-9}
& 5  & 0.15& 0.63& 0.99& 0.69&0.39&0.97&0.56\\
\cline{2-9}
& 6  & 0.16& 0.63& 0.99& 0.69&0.39&0.97&0.56\\
\cline{2-9}
& 7  & 0.15& 0.63& 0.99& 0.69&0.39&0.97&0.56\\
\cline{2-9}
& 8  & 0.16& 0.63& 0.75& 0.53&0.39&0.50&0.44\\
\cline{2-9}
& 9  & 0.15& 0.63& 0.99& 0.69&0.39&0.97&0.56\\
\cline{2-9}
& 10 & 0.15& 0.63& 0.99& 0.69&0.39&0.97&0.56\\
\cline{2-9}
& \textbf{Average} & 0.16& 0.63& 0.96& 0.68&0.39&0.93&0.55\\
\hline
\end{tabular}
\end{table*}
\begin{table*}[ht!]
\centering
\caption{Performance metrics across 10 runs for model instance 1 of Jiménez-Navajas et al. \cite{Jimenez+2025} using a generic prompt with RAG}
\label{tab:model1_airbus-generic-withRAG}
\begin{tabular}{|c|c|c|c|c|c|c|c|c|}
\hline
\multirow{11}{*}{\rotatebox{90}{\textbf{Model Instance 1}}}& \textbf{RUN} & \textbf{CodBLEU}& \textbf{Q-Recall}& \textbf{Q-Precision }& \textbf{Q-F-measure}& \textbf{Recall}& \textbf{Precision }&\textbf{F-measure}\\
\cline{2-9}& 1  & 0.16& 0.61& 1.00& 0.69& 0.38& 1.00&0.55\\
\cline{2-9}& 2  & 0.15& 0.61& 1.00& 0.69& 0.38& 1.00&0.55\\
\cline{2-9}& 3  & 0.15& 0.61& 1.00& 0.69& 0.38& 1.00&0.55\\
\cline{2-9}& 4  & 0.15& 0.62& 1.00& 0.70 & 0.39& 1.00&0.56\\
\cline{2-9}& 5  & 0.15& 0.61& 1.00& 0.69& 0.38& 1.00&0.55\\
\cline{2-9}& 6  & 0.15& 0.61& 1.00& 0.69& 0.38& 1.00&0.55\\
\cline{2-9}& 7  & 0.15& 0.61& 1.00& 0.69& 0.38& 1.00&0.55\\
\cline{2-9}& 8  & 0.16& 0.61& 1.00& 0.69& 0.38& 1.00&0.55\\
\cline{2-9}& 9  & 0.15& 0.62& 1.00& 0.70 & 0.39& 1.00&0.56\\
\cline{2-9}& 10 & 0.15& 0.61& 1.00& 0.69& 0.38& 1.00&0.55\\
\cline{2-9}& \textbf{Average} & 0.16& 0.61& 1.00& 0.69& 0.39& 1.00&0.55\\
\hline
\end{tabular}
\end{table*}

\begin{table*}[ht]
\centering
\caption{Performance metrics across 10 runs for model instance 1 of Jiménez-Navajas et al. \cite{Jimenez+2025} using a specific prompt without RAG}
\label{tab:model1_airbus-specific-withoutRAG}
\begin{tabular}{|c|c|c|c|c|c|c|c|c|}
\hline
\multirow{11}{*}{\rotatebox{90}{\textbf{Model Instance 1 }}}& \textbf{RUN} & \textbf{CodeBLEU}& \textbf{Q-Recall}& \textbf{Q-Precision}& \textbf{Q-F-measure}& \textbf{Recall}& \textbf{Precision}&\textbf{F-measure}\\
\cline{2-9}& 1  & 0.59& 1.00& 1.00& 1.00& 0.43& 1.00&0.60\\
\cline{2-9}& 2  & 0.52& 1.00& 1.00& 1.00& 0.43& 1.00&0.60\\
\cline{2-9}& 3  & 0.57& 0.98& 1.00& 0.99& 0.41& 1.00&0.59\\
\cline{2-9}& 4  & 0.58& 1.00& 1.00& 1.00& 0.43& 1.00&0.60\\
\cline{2-9}& 5  & 0.58& 1.00& 0.99& 0.99& 0.43& 0.97&0.60\\
\cline{2-9}& 6  & 0.58& 1.00& 0.99& 0.99& 0.43& 0.97&0.59\\
\cline{2-9}& 7  & 0.52& 1.00& 1.00& 1.00& 0.43& 1.00&0.60\\
\cline{2-9}& 8  & 0.58& 1.00& 0.99& 0.99& 0.43& 0.97&0.59\\
\cline{2-9}& 9  & 0.58& 0.97& 1.00& 0.98& 0.40& 1.00&0.57\\
\cline{2-9}& 10 & 0.58& 0.97& 1.00& 0.98& 0.40& 1.00&0.57\\
\cline{2-9}& \textbf{Average} & 0.57& 0.99& 1.00& 0.99& 0.42& 0.99&0.59\\
\hline
\end{tabular}
\end{table*}

\begin{table*}[ht!]
\centering
\caption{Performance metrics across 10 runs for model instance 1 of Jiménez-Navajas et al. \cite{Jimenez+2025} using a specific prompt with RAG}
\label{tab:model1_airbus-specific-withRAG}
\begin{tabular}{|c|c|c|c|c|c|c|c|c|}
\hline
\multirow{11}{*}{\rotatebox{90}{\textbf{Model Instance 1 }}}& \textbf{RUN} & \textbf{Code BLEU }& \textbf{Q-Recall}& \textbf{Q-Precision}& \textbf{Q-F-measure}& \textbf{Recall}& \textbf{Precision}&\textbf{F-measure}\\
\cline{2-9}& 1  & 0.59& 1.00& 1.00& 1.00& 0.43& 1.00&0.60\\
\cline{2-9}& 2  & 0.58& 1.00& 1.00& 1.00& 0.43& 1.00&0.60\\
\cline{2-9}& 3  & 0.59& 1.00& 1.00& 1.00& 0.43& 1.00&0.60\\
\cline{2-9}& 4  & 0.57& 0.98& 1.00& 0.99& 0.41& 1.00&0.59\\
\cline{2-9}& 5  & 0.57& 0.98& 1.00& 0.99& 0.41& 1.00&0.59\\
\cline{2-9}& 6  & 0.59& 1.00& 1.00& 1.00& 0.43& 1.00&0.60\\
\cline{2-9}& 7  & 0.57& 0.98& 1.00& 0.99& 0.41& 1.00&0.59\\
\cline{2-9}& 8  & 0.57& 0.98& 1.00& 0.99& 0.41& 1.00&0.59\\
\cline{2-9}& 9  & 0.53& 1.00& 1.00& 1.00& 0.43& 1.00&0.60\\
\cline{2-9}& 10 & 0.60& 1.00& 1.00& 1.00& 0.43& 1.00&0.60\\
\cline{2-9}& \textbf{Average} & 0.58& 0.99& 1.00& 1.00& 0.42& 1.00&0.60\\
\hline
\end{tabular}
\end{table*}

\section{Conclusion and Future Work} \label{conclusion-and-future-work}
In this paper, we have proposed a new research direction with a number of novel ideas and emerging experimental results concerning the first steps towards their validation. We have enabled generating quantum code from UML models by leveraging LLMs, specifically GPT-4o. Unlike traditional methods that have relied on handcrafted transformation rules and deep domain expertise, our approach has shown that using a well-designed, specific prompt enables GPT-4o to produce more accurate results with reduced manual effort. We have also incorporated Qiskit repositories as external contextual knowledge to make the generation process context-aware. However, the results have indicated that this external context has not improved the performance, suggesting that the retrieved information may not have been sufficiently relevant or useful for this specific task.
For future work, we plan to enhance the effectiveness of the RAG pipeline by incorporating more relevant external sources, particularly datasets containing aligned pairs of UML model instances and their corresponding quantum code. Additionally, we aim to improve query formulation, evaluate other LLMs, such as Claude, and extend our analysis using a broader set of evaluation metrics. Most importantly, we will investigate the other ideas proposed in the research questions outlined in Section \ref{proposed-approach}.

\section*{Software and Data Availability}
The source code is available at \url{https://github.com/qas-lab/quantumcodegeneration}. The research data adopted from \cite{Jimenez+2025} are available at \cite{JimenezNavajas+2024}.

\section*{Acknowledgment}
This work is funded by a grant from the Colorado Office of Economic Development and International Trade (OEDIT).

\bibliographystyle{ieeetr}
\bibliography{references}

\end{document}